\documentclass{ws-ijgmmp}

\begin{document}

\markboth{Jun-ichi Note}
{Fourier transform from momentum space to twistor space}

%
\catchline{}{}{}{}{}
%

\title{Fourier transform from momentum space to twistor space
}

\author{Jun-ichi Note
}

\address{Research Institute of Science \& Technology, 
College of Science and 
Technology, \\
Nihon University,
Tokyo 101-8308, Japan\\
\email{note.junichi20@nihon-u.ac.jp
} }

\maketitle


\begin{abstract}
Several methods use the Fourier transform from momentum space to twistor space
to analyze scattering amplitudes in Yang--Mills theory. 
However, the transform has not been defined as a concrete complex
integral 
when the  twistor space is a three-dimensional complex projective space.
To the best of our knowledge, this is the first study to define it 
as well as its inverse in terms of a concrete complex integral. In addition, our study is the first to 
show that the Fourier transform is an isomorphism
from the zeroth \v{C}ech cohomology group
to the first one.
Moreover, the well-known twistor operator representations in twistor theory literature are shown to be valid for the Fourier transform and its inverse transform.
Finally, we identify  functions over which the application of the operators is closed. 
\end{abstract}


\section{Introduction}	
The introduction of twistor string theory by Witten in 2004 \cite{Wit}
triggered the development of  new methods for analyzing
scattering amplitudes in Yang--Mills theory and gravity theory \cite{ElHu}.
In twistor string theory, the tree-level maximally helicity violating (MHV)
amplitudes of four-dimensional $\mathcal{N}=4$ super Yang--Mills 
(SYM) theory are analyzed by applying a Fourier transform from momentum space
to twistor space.
This transform is sometimes referred to as
a half-Fourier transform \cite{BBKR}
because it is a transform with respect to the spinor variable
$\tilde{\pi}_{\alpha}$ alone in the light-like momentum vector
$p_{\alpha \dot{\alpha}}=\tilde{\pi}_{\alpha} \pi_{\dot{\alpha}}$, 
where $\alpha=0,1$ and $\dot{\alpha}=\dot{0},\dot{1}$.
It has been shown that a gluon MHV amplitude in twistor space is
supported on an algebraic curve of degree one.
This amplitude is deduced from the string model whose target
space is the twistor space.
In general, it is conjectured that an $l$-loop amplitude with
$q$ gluons of negative helicity in twistor space is supported on 
an algebraic curve
 of degree $d=q-1+l$.
Another example that demonstrates
 the utility of the 
Fourier transform
is the translation of the Britto--Cachazo--Feng--Witten (BCFW) recursion relations \cite{BCF,BCFW} 
concerning scattering amplitudes in momentum space.
This shows that scattering amplitudes in twistor space
have the same recursion relations \cite{MS2,ACCK2}.
In addition, the Fourier transform is also used in
the representation of 
the $n$-particle (next-to)$^k$ maximally helicity violating (N$^k$MHV)
amplitude 
, i.e., the amplitude with $k+2$ negative helicity gluons and
$n-k-2$ positive helicity gluons,
in planar $\mathcal{N}=4$ SYM  theory as an integral on a
Grassmann manifold with the aid of twistors \cite{ACCK,MS}.
This makes the symmetries of the N$^{k}$MHV amplitude, i.e.,
the cyclic symmetry of the external particles, the superconformal symmetry,
and the Yangian symmetry, manifest.
Furthermore, 
relativistic particles of various helicities can be analyzed similarly 
in terms of  twistor variables with the aid of the Fourier
transform \cite{DNOS}.

When the spacetime metric signature is $(2,2)$, i.e., the ultra-hyperbolic 
spacetime,
the spinors $\pi_{\dot{\alpha}}$
and $\tilde{\pi}_{\alpha}$ of the light-like momentum vector are real, and
the corresponding twistor space is the three-dimensional real
projective space $\mathbb{RP}^{3}$ \cite{Wit}.
Thus, the Fourier transform from momentum space to twistor space is
the well-known one from the standard analysis.
However, when the spacetime metric signature is $(1,3)$, i.e.,
the  Minkowski space $\mathbb{M}$,
the spinor $\pi_{\dot{\alpha}}$ is complex and the spinor $\tilde{\pi}_{\alpha}$
is its complex conjugate, i.e.,
$\tilde{\pi}_{\alpha} = \bar{\pi}_{\alpha}$, and
the corresponding twistor space is the three-dimensional 
complex projective space $\mathbb{CP}^{3}$ \cite{Pen1}.
In this case, the Fourier transform is a complex 
integral with respect to
the spinor variable $\bar{\pi}_{\alpha}$.
Hence, the definition is different from that in the real-number case.
Thus far, this Fourier transform has not been defined as a concrete
complex integral \cite{ACCK,DNOS}. Therefore, 
the concrete integration method, 
the functions to which  the integral can be applied,
and the explicit results of the integral  
 are  all unknown.
Investigating these issues could facilitate studies of the mathematical
structures of the scattering amplitudes in $\mathcal{N}=4$ SYM theory as well as
those of relativistic particles, which remain unknown.

This paper proposes a new definition of 
the Fourier transform as a concrete complex integral
of functions in momentum space in the following way.
First, we assume that the momentum-space functions $f(\bar{\pi}_{\alpha},
\pi_{\dot{\alpha}})$ can be expanded 
in terms of positive powers
$\bar{\pi}_{\alpha}$, 
$(\bar{\pi}_{0})^{a}(\bar{\pi}_{1})^{b}$ ($a,b=0,1,2,\cdots$).
Here, we denote the momentum-space coordinates
by $(\bar{\pi}_{\alpha}, \pi_{\dot{\alpha}})$.
Then, we propose the Fourier transform of $f(\bar{\pi}_{\alpha},\pi_{\dot{\alpha}})$
as a concrete complex integral with respect to the variables $\bar{\pi}_{\alpha}$
such that it transforms $f(\bar{\pi}_{\alpha},\pi_{\dot{\alpha}})$ to
a function of twistor-space coordinates
 $(\omega^{\alpha},\pi_{\dot{\alpha}})$, where we define
the spinor $\omega^{\alpha}$ by contracting 
$\pi_{\dot{\alpha}}$ with the coordinates 
$(x^{\alpha \dot{\alpha}})$ of $\mathbb{M}$, i.e.,  
$\omega^{\alpha} := ix^{\alpha \dot{\alpha}} \pi_{\dot{\alpha}}$.  
Here, we determine the integral contour
such that  
$(\bar{\pi}_{0})^{0}(\bar{\pi}_{1})^{0}=1$ is transformed to a complex function
that represents the delta function  
of Sato's hyperfunction.
In this definition, the Fourier transform of 
$f(\bar{\pi}_{\alpha},\pi_{\dot{\alpha}})$ is
the twistor-space function
 $\tilde{f}(\omega^{\alpha},\pi_{\dot{\alpha}})$
 expanded in terms of negative powers $\omega^{\alpha}$,
$(\omega^{0})^{-a-1}(\omega^{1})^{-b-1}$.
Next, we propose an inverse Fourier transform of 
twistor-space functions $\tilde{f}(\omega^{\alpha},\pi_{\dot{\alpha}})$
to momentum space.
We define it as a concrete complex integral that transforms a
complex function on twistor space that represents
the delta function of Sato's hyperfunction to $1$.
In this definition, the momentum-space function $f(\bar{\pi}_{\alpha},
\pi_{\dot{\alpha}})$ is obtained from the inverse Fourier transform of
the function $\tilde{f}(\omega^{\alpha},\pi_{\dot{\alpha}})$, which is
the Fourier transform of $f(\bar{\pi}_{\alpha},\pi_{\dot{\alpha}})$.
Furthermore, the  twistor-space function  
$\tilde{f}(\omega^{\alpha},\pi_{\dot{\alpha}})$ is obtained from the Fourier
transform of the function $f(\bar{\pi}_{\alpha},\pi_{\dot{\alpha}})$, 
which is the inverse Fourier
transform of $\tilde{f}(\omega^{\alpha},\pi_{\dot{\alpha}})$.
In other words, this
inverse Fourier transform is precisely the inverse map
of the Fourier transform.

From the perspective of the \v{C}ech cohomology group, 
the power functions
$(\bar{\pi}_{0})^{a}(\bar{\pi}_{1})^{b}$ ($a+b=n$, $n=0,1,2,\cdots$) to 
which the Fourier transform is applied form
a basis for the zeroth \v{C}ech
cohomology group on $\mathbb{CP}^{1}$ with coefficients in the sheaf
$\mathcal{O}(n)$ consisting of homogeneous functions of degree $n$.
Furthermore, 
the power functions $(\omega^{0})^{-a-1}(\omega^{1})^{-b-1}$, 
which are obtained from the Fourier transform of
$(\bar{\pi}_{0})^{a}(\bar{\pi}_{1})^{b}$, form a basis for the first \v{C}ech 
cohomology group on $\mathbb{CP}^{1}$ with coefficients in the sheaf
$\mathcal{O}(-n-2)$ consisting of homogeneous functions of degree $-n-2$.
Accordingly, this study is the first to show that
the Fourier transform is an isomorphism from the zeroth \v{C}ech
cohomology group to the first one, and  
the momentum-space functions to which 
the Fourier transform is applied are direct sums of the tensor product of 
the two zeroth \v{C}ech cohomology groups. 
Furthermore, the  twistor-space functions
obtained from the Fourier transform are shown to be direct sums of the tensor product 
of the zeroth and first
 \v{C}ech cohomology groups.

In the canonical quantization procedure in twistor theory, the variables
$\omega^{\alpha}$ and $\bar{\pi}_{\alpha}$ are replaced by their corresponding
operators \cite{Pen5,PM,DN}.
As in standard quantum mechanics, one of these 
operator types is represented as a multiplication operator, and the other 
is represented as a partial differential operator.
We show that such operator representations are also valid for
the Fourier transform and its inverse.
Finally, we prove that the application of these
operators is closed in the momentum-space functions
and twistor-space functions, respectively.

This paper is organized as follows. Sec. \ref{AdFt}
proposes a new definition of the
Fourier transform from momentum space to twistor space
as well as its inverse.
Sec. \ref{Ci} describes the functions to which  
the (inverse) Fourier transform is applied 
in terms of \v{C}ech cohomology groups.
Sec. \ref{Trotto} shows that the well-known operator representations 
in the  twistor theory literature are also valid for the (inverse)
Fourier transform.
Finally, Sec. \ref{Sd} summarizes our study.

\section{New definition of the Fourier transform}
\label{AdFt}

\subsection{Brief review of the delta function of two variables}

The delta function of two variables can be defined as
a product of single-variable delta functions \cite{Ka}:
\begin{align}
\delta (x_{1}, x_{2}) 
&:= \delta (x_{1}) \delta (x_{2}) \notag \\
&= \prod_{j=1}^{2} \dfrac{1}{-2 \pi i}
\left( \dfrac{1}{x_{j}+i0} - \dfrac{1}{x_{j}-i0} \right) \notag \\
&= \dfrac{1}{(-2 \pi i)^{2}} \sum_{\sigma} 
\dfrac{\text{sgn} \sigma}{(x_{1}+i\sigma_{1}0)(x_{2}+i\sigma_{2}0)},
\label{B1}
\end{align}
where $\sigma = (\sigma_{1}, \sigma_{2})$, $\sigma_{1},\sigma_{2}=\pm 1$, and 
$\textrm{sgn} \sigma = \sigma_{1} \sigma_{2}$.
By defining  the $\sigma$ quadrant as
\begin{align}
\varGamma_{\sigma} = \left\{\,(\eta_{1}, \eta_{2}) \in \mathbb{R}^{2} \,|\,
\sigma_{1} \eta_{1} >0,\, \sigma_{2} \eta_{2} > 0
\,\right\}
,
\label{2}
\end{align}
each term in Eq. $(\ref{B1})$ can be interpreted as the 
boundary value of the two-variable complex function
\begin{align}
\dfrac{1}{(-2 \pi i)^{2}} \dfrac{\textrm{sgn} \sigma}{z_{1} z_{2}}
\label{1}
\end{align}
at $\mathbb{R}^{2} + i\varGamma_{(1,1)}$, 
$\mathbb{R}^{2} + i\varGamma_{(1,-1)}$, $\mathbb{R}^{2} + i\varGamma_{(-1,1)}$,
and $\mathbb{R}^{2} + i\varGamma_{(-1,-1)}$.

In sections \ref{2-2} and \ref{2-3}, we propose a new
definition of the Fourier transform and its inverse
transform by use of Eq. $(\ref{1})$.

\subsection{Fourier transform}
\label{2-2}

To derive a new definition of the Fourier transform as a concrete complex integral, 
we focus on momentum-space functions, which
have the power series expansion
\begin{align}
f(\bar{\pi}_{\alpha}, \pi_{\dot{\alpha}})
= \sum_{a,b=0}^{\infty} C_{ab} (\pi_{\dot{0}}, \pi_{\dot{1}})
(\bar{\pi}_{0})^{a} (\bar{\pi}_{1})^{b}.
\label{3}
\end{align}
Here, $C_{ab}(\pi_{\dot{0}}, \pi_{\dot{1}})$ denotes a holomorphic function
of the spinor $\pi_{\dot{\alpha}}$.
To transform such functions $f(\bar{\pi}_{\alpha},\pi_{\dot{\alpha}})$
to twistor space, we propose the Fourier transform 
from the power series terms
$(\bar{\pi}_{0})^{a}(\bar{\pi}_{1})^{b}$ ($a,b=0,1,2,\cdots$) to functions
of $\omega^{\alpha}$.

\begin{definition}
First, we propose the Fourier transform of
$(\bar{\pi}_{0})^{0}(\bar{\pi}_{1})^{0}=1$ as follows:
\begin{align}
\mathcal{F}[1] 
&:= \dfrac{1}{(-2 \pi i)^{2}} \int_{-i\varGamma_{\sigma}}
e^{-\omega^{0} \bar{\pi}_{0} -\omega^{1} \bar{\pi}_{1}} d \bar{\pi}_{0}\,
d \bar{\pi}_{1}
\notag \\
&= \dfrac{1}{(-2\pi i)^{2}} \dfrac{\textrm{sgn} \sigma}{\omega^{0} \omega^{1}}, \quad
(\omega^{0}, \omega^{1}) \in \mathbb{R}^{2} + i \varGamma_{\sigma}.
\label{4}
\end{align}
This is a complex function similar to Eq. ($\ref{1}$), which 
represents the delta function.
\end{definition}

When we take $\sigma = (1,-1)$, the integral in
Eq. ($\ref{4}$) represents an improper integral from $-\infty$ to $0$
on the imaginary axis for  the variable $\bar{\pi}_{0}$ and 
from $0$ to $\infty$ on the imaginary axis for the variable $\bar{\pi}_{1}$:  
\begin{align}
&\dfrac{1}{(-2 \pi i)^{2}} \int_{-i \varGamma_{(1,-1)}}
e^{-\omega^{0} \bar{\pi}_{0} - \omega^{1} \bar{\pi}_{1}} d\bar{\pi}_{0}\,
d\bar{\pi}_{1}
\notag \\
&= \dfrac{1}{(-2 \pi i)^{2}} \int_{-\infty}^{0} 
e^{-\omega^{0} (i\text{Im} \bar{\pi}_{0})} d(i \text{Im} \bar{\pi}_{0})
\int_{0}^{\infty} 
e^{-\omega^{1} (i\text{Im} \bar{\pi}_{1})} d(i \text{Im} \bar{\pi}_{1})
\notag \\
&= \dfrac{1}{(-2 \pi i)^{2}} \dfrac{-1}{\omega^{0} \omega^{1}}.
\label{5}
\end{align}
This is uniformly convergent on $(\omega^{0}, \omega^{1})
\in \mathbb{R}^{2} + i\varGamma_{(1,-1)}$. For all other $\sigma$,
the meaning of the integral sign in Eq. ($\ref{4}$) is the same,
and the integral is uniformly convergent on $(\omega^{0}, \omega^{1})
\in \mathbb{R}^{2} + i\varGamma_{\sigma}$.

\begin{theorem}
The Fourier transform
of the momentum-space function $f(\bar{\pi}_{\alpha}, \pi_{\dot{\alpha}})$ 
is
\begin{align}
&\mathcal{F}\left[ f(\bar{\pi}_{\alpha}, \pi_{\dot{\alpha}}) \right]
\notag \\
&:= \dfrac{1}{(-2 \pi i)^{2}} \int_{-i \varGamma_{\sigma}} 
f(\bar{\pi}_{\alpha}, \pi_{\dot{\alpha}})\,
e^{-\omega^{0} \bar{\pi}_{0} - \omega^{1} \bar{\pi}_{1}}\,
d\bar{\pi}_{0} \, d\bar{\pi}_{1}
\notag \\
&= \sum_{a,b=0}^{\infty} C_{ab} (\pi_{\dot{0}}, \pi_{\dot{1}})
\dfrac{\textrm{sgn} \sigma}{(-2 \pi i)^{2}} 
\dfrac{\varGamma (a+1) \varGamma (b+1)}
{(\omega^{0})^{a+1} (\omega^{1})^{b+1}}, \quad
(\omega^{0}, \omega^{1}) \in \mathbb{R}^{2} + i \varGamma_{\sigma}.
\label{6}
\end{align}
By using the analytic continuation method, we can extend the  
$\{(\omega^{0}, \omega^{1})\}$ region from 
$\mathbb{R}^{2} + i \varGamma_{\sigma}$ to the direct product of
$\mathbb{C}^{*} = \{\omega^{0} \in \mathbb{C} \,|\, \omega^{0} \ne 0\}$
and 
$\mathbb{C}^{*} = \{\omega^{1} \in \mathbb{C} \,|\, \omega^{1} \ne 0\}$.
Hereafter, we consider the $\{(\omega^{0},\omega^{1})\}$ region
in Eq. ($\ref{6}$) as $\mathbb{C}^{*} \times \mathbb{C}^{*}$.
\end{theorem}

\begin{proof}
From Eq. ($\ref{4}$), the Fourier transform of the power series term
$(\bar{\pi}_{0})^{a} (\bar{\pi}_{1})^{b}$ ($a,b=0,1,2,\cdots$) is
\begin{align}
\mathcal{F} \left[ (\bar{\pi}_{0})^{a} (\bar{\pi}_{1})^{b} \right]
&:= \dfrac{1}{(-2 \pi i)^{2}} \int_{-i\varGamma_{\sigma}}
(\bar{\pi}_{0})^{a} (\bar{\pi}_{1})^{b} 
e^{-\omega^{0} \bar{\pi}_{0} -\omega^{1} \bar{\pi}_{1}}
d\bar{\pi}_{0} \, d\bar{\pi}_{1}
\notag \\
&= \dfrac{\textrm{sgn} \sigma}{(-2 \pi i)^{2}} 
\dfrac{\varGamma (a+1) \varGamma (b+1)}
{(\omega^{0})^{a+1} (\omega^{1})^{b+1}}
, \quad (\omega^{0}, \omega^{1}) \in \mathbb{R}^{2} + i \varGamma_{\sigma}.
\label{7}
\end{align}
The integral in Eq. (\ref{7}) can be calculated in
the same way as that in Eq. (\ref{4}).
Here, the integration formula  
\begin{align}
\int_{0}^{\infty} x^{z-1} e^{-wx} dx  = \dfrac{\varGamma (z)}{w^{z}}, \quad
\text{Re}\, w > 0, \quad \text{Re}\,z > 0
\label{8}
\end{align}
has been used.
Thus, from Eqs. ($\ref{3}$) and ($\ref{7}$),
Eq. ($\ref{6}$) is obtained.
\end{proof}

\subsection{Inverse Fourier transform}
\label{2-3}

Now, we propose the inverse Fourier transform from twistor space to
momentum space.
Here, we focus on twistor-space functions $\tilde{f}
(\omega^{\alpha}, \pi_{\dot{\alpha}})$ that can be expanded in terms of the
power series terms
$(\omega^{0})^{-a-1} (\omega^{1})^{-b-1}$ ($a,b=0,1,2,\cdots$), 
similar to Eq. $(\ref{6})$:
\begin{align} 
\tilde{f}(\omega^{\alpha}, \pi_{\dot{\alpha}})
= \sum_{a,b=0}^{\infty} \dfrac{D_{ab}(\pi_{\dot{0}},\pi_{\dot{1}})}
{(\omega^{0})^{a+1} (\omega^{1})^{b+1}}.
\label{9}
\end{align}
Here, $D_{ab}(\pi_{\dot{0}}, \pi_{\dot{1}})$ is a holomorphic function of
$\pi_{\dot{0}}$ and $\pi_{\dot{1}}$.

\begin{definition}
We propose that the inverse Fourier transform of the complex 
function (similar to Eq. ($\ref{1}$))
\begin{align}
\dfrac{1}{(-2 \pi i)^{2}} \dfrac{\textrm{sgn} \sigma}{\omega^{0} \omega^{1}}
\label{10}
\end{align} 
is equal to 1. In other words,
we propose that the inverse Fourier transform of 
$1/(\omega^{0} \omega^{1})$, $\mathcal{F}^{-1}[1/(\omega^{0} \omega^{1})]$
is equal to $(-2 \pi i)^{2} \textrm{sgn} \sigma$
because $(\textrm{sgn} \sigma)^{2}=1$:
\begin{align}
\mathcal{F}^{-1} \left[ \dfrac{1}{\omega^{0} \omega^{1}} \right]
&:= \lim_{\varepsilon \rightarrow 0} (-2 \pi i)^{2} \varepsilon^{2}
\int_{\varGamma_{\sigma}} (\omega^{0})^{-1+\varepsilon}
(\omega^{1})^{-1+\varepsilon} 
e^{\omega^{0} \bar{\pi}_{0} + \omega^{1} \bar{\pi}_{1}} d \omega^{0}\,
d \omega^{1}
\notag \\
&= (-2 \pi i)^{2} \textrm{sgn} \sigma, \quad (\bar{\pi}_{0}, \bar{\pi}_{1}) \in
- \varGamma_{\sigma} + i \mathbb{R}^{2}.
\label{11} 
\end{align}
\end{definition}

When we take $\sigma = (1,-1)$, the integral in
 Eq. ($\ref{11}$) represents an improper integral from $0$ to $\infty$
on the real axis for  the variable $\omega^{0}$ and 
from $-\infty$ to $0$ on the real axis for the variable
$\omega^{1}$:
\begin{align}
&\lim_{\varepsilon \rightarrow 0} (-2 \pi i)^{2} \varepsilon^{2}
\int_{\varGamma_{(1,-1)}} (\omega^{0})^{-1+\varepsilon} 
(\omega^{1})^{-1+\varepsilon} 
e^{\omega^{0} \bar{\pi}_{0} + \omega^{1} \bar{\pi}_{1}} 
d \omega^{0} \, d\omega^{1}
\notag \\
= 
&\lim_{\varepsilon \rightarrow 0} (-2 \pi i)^{2} \varepsilon^{2}
\int_{0}^{\infty} \left( \text{Re} \omega^{0} \right)^{-1+\varepsilon}
e^{(\text{Re} \omega^{0}) \bar{\pi}_{0}} d \left( \text{Re} \omega^{0} \right)
\notag \\
&  
\times
\int_{-\infty}^{0} \left( \text{Re} \omega^{1} \right)^{-1+\varepsilon}
e^{(\text{Re} \omega^{1}) \bar{\pi}_{1}} d \left( \text{Re} \omega^{1} \right)
\notag \\
= 
&(-2 \pi i)^{2} (-1), \quad \left( \bar{\pi}_{0}, \bar{\pi}_{1} \right)
\in - \varGamma_{(1,-1)} + i \mathbb{R}^{2}. 
\label{12}
\end{align}
Here, the integration formula
\begin{align} 
\int_{0}^{\infty} x^{z-1+\varepsilon} e^{-wx} dx = 
\dfrac{\varGamma ( z+\varepsilon )}{w^{z+\varepsilon}}, \quad
\text{Re} w > 0
\label{13}
\end{align}
and the  gamma function formula 
\begin{align} 
\varGamma (\varepsilon) = \dfrac{1}{\varepsilon} - \gamma
+ \mathcal{O} (\varepsilon)
\label{14}
\end{align}
have been used, where $\gamma$ is the Euler--Mascheroni constant.
For all other $\sigma$, the meaning of the integral sign in Eq. $(\ref{11})$
is the same, and the integral is uniformly convergent on
$(\bar{\pi}_{0}, \bar{\pi}_{1}) \in - \varGamma_{\sigma}
+i \mathbb{R}^{2}$.

\begin{theorem}
The inverse
Fourier transform of the twistor-space function 
$\tilde{f}(\omega^{\alpha}, \pi_{\dot{\alpha}})$ is
\begin{align}
&\mathcal{F}^{-1} \left[ \tilde{f}(\omega^{\alpha}, \pi_{\dot{\alpha}}) \right]
\notag \\
&:= \lim_{\varepsilon \rightarrow 0} (-2 \pi i)^{2} \varepsilon^{2}
\int_{\varGamma_{\sigma}} \tilde{f}(\omega^{\alpha}, \pi_{\dot{\alpha}})
(\omega^{0})^{\varepsilon} (\omega^{1})^{\varepsilon}
e^{\omega^{0} \bar{\pi}_{0} + \omega^{1} \bar{\pi}_{1}}
d\omega^{0} \, d\omega^{1}
\notag \\
&= \sum_{a,b=0}^{\infty} D_{ab}(\pi_{\dot{0}}, \pi_{\dot{1}}) (-2 \pi i)^{2}
\textrm{sgn} \sigma \dfrac{(\bar{\pi}_{0})^{a} (\bar{\pi}_{1})^{b}}
{\varGamma (a+1) \varGamma (b+1)}, 
\notag \\
&
(\bar{\pi}_{0}, \bar{\pi}_{1}) \in - \varGamma_{\sigma} + i \mathbb{R}^{2}.
\label{15}
\end{align}
By using the  analytic continuation method, we can extend the  
$\{(\bar{\pi}_{0}, \bar{\pi}_{1})\}$ region
from $-\varGamma_{\sigma}
+ i \mathbb{R}^{2}$ to $\mathbb{C}^{2}$.
Hereafter, we consider the $\{(\bar{\pi}_{0}, \bar{\pi}_{1})\}$ region
in Eq. $(\ref{15})$ as $\mathbb{C}^{2}$.
\end{theorem}

\begin{proof}
From Eq. (\ref{11}), the inverse Fourier transform of the power series term
$(\omega^{0})^{-a-1} (\omega^{1})^{-b-1}$ ($a,b=0,1,2,\cdots$)
is
\begin{align}
&\mathcal{F}^{-1} \left[ (\omega^{0})^{-a-1} (\omega^{1})^{-b-1} \right]
\notag \\
&:= \lim_{\varepsilon \rightarrow 0} (-2 \pi i)^{2} \varepsilon^{2}
\int_{\varGamma_{\sigma}} (\omega^{0})^{-a-1+\varepsilon}
(\omega^{1})^{-b-1+\varepsilon} 
e^{\omega^{0} \bar{\pi}_{0} + \omega^{1} \bar{\pi}_{1}} 
d\omega^{0} \,d\omega^{1}
\notag \\
&= (-2 \pi i)^{2} \textrm{sgn} \sigma \dfrac{(\bar{\pi}_{0})^{a} (\bar{\pi}_{1})^{b}}
{\varGamma (a+1) \varGamma (b+1)}, \quad
(\bar{\pi}_{0}, \bar{\pi}_{1}) \in - \varGamma_{\sigma} + i \mathbb{R}^{2}.
\label{16}
\end{align}
The integral in
Eq. ($\ref{16}$) can be calculated in the same way as that in
Eq. ($\ref{11}$).
Here, we have used
the integration formula in Eq. ($\ref{13}$) and
the gamma function formula 
\begin{align}
\varGamma (-n+\varepsilon ) = \dfrac{(-1)^{n}}{n!} \left[
\dfrac{1}{\varepsilon} + \psi_{1} (n+1) + \mathcal{O}(\varepsilon ) \right], \quad
n=0,1,2,\cdots,
\label{17}
\end{align}
where $\psi_{1}(n+1) := \varGamma^{\prime} (n+1) / \varGamma(n+1)
= \sum_{p=1}^{n} p^{-1} - \gamma$.
Thus, from Eqs. ($\ref{9}$) and ($\ref{16}$),
Eq. ($\ref{15}$) is obtained.
\end{proof}

\begin{corollary}
The inverse Fourier transform in Eq. $(\ref{15})$
is precisely the inverse map of the Fourier transform
in Eq. $(\ref{6})$, i.e.,
\begin{align}
\mathcal{F}^{-1} \mathcal{F} 
\left[ f(\bar{\pi}_{\alpha}, \pi_{\dot{\alpha}}) \right]
= f(\bar{\pi}_{\alpha}, \pi_{\dot{\alpha}}), \quad
\mathcal{F} \mathcal{F}^{-1} 
\left[ \tilde{f}(\omega^{\alpha}, \pi_{\dot{\alpha}}) \right]
= \tilde{f}(\omega^{\alpha}, \pi_{\dot{\alpha}}).
\label{18}
\end{align}
\end{corollary}

\begin{proof}
From Eqs. $(\ref{7})$ and $(\ref{16})$, we have
\begin{align} 
\mathcal{F}^{-1} \mathcal{F} \left[ (\bar{\pi}_{0})^{a} (\bar{\pi}_{1})^{b} \right]
&= \mathcal{F}^{-1} \left[ \dfrac{\textrm{sgn} \sigma}{(-2 \pi i)^{2}}
\dfrac{\varGamma(a+1) \varGamma(b+1)}
{(\omega^{0})^{a+1} (\omega^{1})^{b+1}} \right]
\notag \\
&= (\bar{\pi}_{0})^{a} (\bar{\pi}_{1})^{b},
\label{19} 
\end{align}
\begin{align}
\mathcal{F} \mathcal{F}^{-1} 
\left[(\omega^{0})^{-a-1} (\omega^{1})^{-b-1}\right]
&= \mathcal{F} \left[ (-2 \pi i)^{2} \textrm{sgn} \sigma 
\dfrac{(\bar{\pi}_{0})^{a} (\bar{\pi}_{1})^{b}}{\varGamma(a+1) \varGamma(b+1)}
\right]
\notag \\
&= (\omega^{0})^{-a-1} (\omega^{1})^{-b-1}.
\label{20}
\end{align}
Thus, from Eqs. $(\ref{3})$, $(\ref{9})$, $(\ref{19})$, and $(\ref{20})$,
 we obtain
 Eq. $(\ref{18})$. 
\end{proof}

\section{Cohomological interpretation}
\label{Ci}

\subsection{Brief review of \v{C}ech cohomology group}

We now consider the one-dimensional complex projective space
$\mathbb{CP}^{1}$ covered by the two open sets
\begin{align}
U_{0} = \left\{ 
\left( \bar{\pi}_{0}, \bar{\pi}_{1} \right) \in \mathbb{CP}^{1}
\bigl| \bar{\pi}_{0} \ne 0 \right\}, \quad
U_{1} = \left\{ 
\left( \bar{\pi}_{0}, \bar{\pi}_{1} \right) \in \mathbb{CP}^{1}
\bigl| \bar{\pi}_{1} \ne 0 \right\}.
\label{21}
\end{align}
Let $\mathcal{O}(n)$ ($n \in \mathbb{Z}$)
be the sheaf of germs of homogeneous holomorphic functions on
$\mathbb{CP}^{1}$  of degree $n$.
Then, the zeroth \v{C}ech cohomology group on $\mathbb{CP}^{1}$
with coefficients in the sheaf $\mathcal{O}(n)$ is represented as \cite{HT}
\begin{align}
&H^{0} \left( \mathbb{CP}^{1}, \mathcal{O}(n) \right)
\notag \\
&= \left\{ \sum_{r=1}^{n+1} a_{r} \left( \bar{\pi}_{0} \right)^{r-1}
\left( \bar{\pi}_{1} \right)^{n+1-r} 
\Bigg|
a_{r} \in \mathbb{C}
\right\}, \quad
n=0,1,2,\cdots.
\label{22}
\end{align}
This vanishes for negative integers $n=-1,-2,-3,\cdots$.

When $\mathbb{CP}^{1}$ is covered by the open sets
\begin{align}
U^{\prime}_{0} = \left\{ \left( \omega^{0}, \omega^{1} \right) 
\in \mathbb{CP}^{1}
\bigl| \omega^{0} \ne 0 \right\}, \quad
U^{\prime}_{1} = \left\{ \left( \omega^{0}, \omega^{1} \right) 
\in \mathbb{CP}^{1}
\bigl| \omega^{1} \ne 0 \right\},
\label{23}
\end{align}
the first \v{C}ech cohomology group on $\mathbb{CP}^{1}$ with coefficients
in the sheaf $\mathcal{O}(-n-2)$ is represented as \cite{HT}
\begin{align}
&H^{1} \left( \mathbb{CP}^{1}, \mathcal{O}(-n-2) \right)
\notag \\
&=\left\{ \sum_{r=1}^{n+1} \dfrac{a_{r}}{(\omega^{0})^{r} (\omega^{1})^{n+2-r}} 
\Bigg| a_{r} \in \mathbb{C} \right\}, \quad
n=0,1,2,\cdots.
\label{24}
\end{align}
This also vanishes for negative integers $n=-1,-2,-3,\cdots$.
The elements of $H^{1}\left(\mathbb{CP}^{1}, \mathcal{O}(-n-2) \right)$
are equivalent to elements with extra homogeneous holomorphic functions
on $U^{\prime}_{i}$ ($i=0,1$) of degree $(-n-2)$ \cite{HT}: 
\begin{align}
\sum_{r=1}^{n+1} \dfrac{a_{r}}{(\omega^{0})^{r} (\omega^{1})^{n+2-r}}
\sim  
& \sum_{r=1}^{n+1} \dfrac{a_{r}}{(\omega^{0})^{r} (\omega^{1})^{n+2-r}}
\notag \\
&+ \sum_{s=0}^{\infty} \left( 
b_{s} \dfrac{(\omega^{1})^{s}}{(\omega^{0})^{n+2+s}} 
+ c_{s} \dfrac{(\omega^{0})^{s}}{(\omega^{1})^{n+2+s}} 
\right).
\end{align}
When $n=0,1,2,\cdots$, the \v{C}ech cohomology groups in Eqs.
$(\ref{22})$ and $(\ref{24})$ are isomorphic to $\mathbb{C}^{n+1}$. 
However, they are not the same; instead, they are dual spaces of
each other.
This is a particular instance of Serre duality \cite{Koba}.

Now, let $L$ be a complex projective line in the twistor space 
$\mathbb{PT}$ defined as 
$L=\left\{(\omega^{\alpha},\pi_{\dot{\alpha}}) \in \mathbb{PT}
\,|\, \omega^{0} = \omega^{1} = 0\right\} \simeq \mathbb{CP}^{1}$.
Let $\mathbb{PT}-L$ be covered by
\begin{align}
V_{0} = \left\{ (\omega^{\alpha},\pi_{\dot{\alpha}}) \in \mathbb{PT} \bigl|
\omega^{0} \ne 0 \right\}, \quad
V_{1} = \left\{ (\omega^{\alpha},\pi_{\dot{\alpha}}) \in \mathbb{PT} \bigl|
\omega^{1} \ne 0 \right\}.
\label{28}
\end{align}
The first \v{C}ech cohomology
group on $\mathbb{PT}-L$ with coefficients in the sheaf 
$\mathcal{O}(m)$ ($m \in \mathbb{Z}$) is represented as \cite{EH}   
\begin{align}
H^{1}\left( \mathbb{PT}-L, \mathcal{O}(m) \right)
= \left\{ \sum_{j,k>0} \dfrac{A_{jk}(\pi_{\dot{0}},\pi_{\dot{1}})}
{(\omega^{0})^{j} (\omega^{1})^{k}} \right\}.
\label{29}
\end{align}
Here, $A_{jk}(\pi_{\dot{0}},\pi_{\dot{1}})$ denotes the homogeneous
holomorphic function
on $L$  of degree $(j+k+m)$:
$A_{jk}(\pi_{\dot{0}},\pi_{\dot{1}}) \in \varGamma(L,\mathcal{O}(j+k+m))
= H^{0}(L,\mathcal{O}(j+k+m))$.
Hence, the sum in Eq. $(\ref{29})$
is taken for $j>0$ and $k>0$ satisfying $j+k+m \ge 0$,
because $H^{0}(L,\mathcal{O}(j+k+m))$ is $0$ for $j+k+m < 0$.

We emphasize that
Eq. $(\ref{29})$ is precisely the  twistor-space functions to which
the inverse Fourier transform is applied, as seen from Eq. $(\ref{9})$.
In the twistor theory literature, the elements of Eq. $(\ref{29})$ are 
referred to as elementary states \cite{BE}, and they are used in the
calculation of twistor diagrams \cite{PM}.

\subsection{New theorem and its corollary}

Hereafter, we identify the properties of the Fourier
transform in terms of \v{C}ech cohomology groups.

\begin{theorem}
The Fourier transform is an
isomorphism from $H^{0}(\mathbb{CP}^{1}, \mathcal{O}(n))$ to
$H^{1}(\mathbb{CP}^{1},\mathcal{O}(-n-2))$, and
the inverse Fourier transform is an isomorphism 
from $H^{1}(\mathbb{CP}^{1},\mathcal{O}(-n-2))$ to 
$H^{0}(\mathbb{CP}^{1},\mathcal{O}(n))$.
\end{theorem}

\begin{proof}
From Eq. $(\ref{7})$, the Fourier transform of the basis
$\left\{ (\bar{\pi}_{0})^{r-1} (\bar{\pi}_{1})^{n+1-r} \right\}$ of $H^{0}(\mathbb{CP}^{1},\mathcal{O}(n))$
is
\begin{align}
\mathcal{F}\left[ (\bar{\pi}_{0})^{r-1} (\bar{\pi}_{1})^{n+1-r} \right]
= \dfrac{\textrm{sgn} \sigma}{(-2\pi i)^{2}}
\dfrac{\varGamma(r) \varGamma(n+2-r)}
{(\omega^{0})^{r} (\omega^{1})^{n+2-r}}.
\label{25}
\end{align}
Equation $(\ref{25})$ is precisely the basis $\{(\omega^{0})^{-r} (\omega^{1})^{-n-2+r}\}$
of $H^{1}(\mathbb{CP}^{1}, \mathcal{O}(-n-2))$. 
Furthermore, from Eq. $(\ref{16})$, the inverse Fourier transform of
the basis 
$\left\{(\omega^{0})^{-r} (\omega^{1})^{-n-2+r}\right\}$ of  
$H^{1}(\mathbb{CP}^{1}, \mathcal{O}(-n-2))$
is
\begin{align}
\mathcal{F}^{-1} \left[ (\omega^{0})^{-r} (\omega^{1})^{-n-2+r} \right]
= (-2 \pi i)^{2} \textrm{sgn} \sigma
 \dfrac{(\bar{\pi}_{0})^{r-1} (\bar{\pi}_{1})^{n+1-r}}
{\varGamma(r) \varGamma(n+2-r)}.
\label{26}
\end{align}
In addition, if we take
 the inverse Fourier transform of a homogeneous holomorphic function on
$U^{\prime}_{i}$ ($i=0,1$)  of degree $(-n-2)$,
which is added to the equivalence class of 
$H^{1}(\mathbb{CP}^{1}, \mathcal{O}(-n-2))$, i.e., 
$(\omega^{1})^{s}/(\omega^{0})^{n+2+s}$ or
$(\omega^{0})^{s}/(\omega^{1})^{n+2+s}$ ($s=0,1,2,\cdots$), we obtain $0$
because the gamma function in the denominator diverges:
\begin{align}
\mathcal{F}^{-1} \left[(\omega^{1})^{s}/(\omega^{0})^{n+2+s} \right]
= \mathcal{F}^{-1} \left[(\omega^{0})^{s}/(\omega^{1})^{n+2+s} \right] = 0.
\label{27}
\end{align}
Therefore, the inverse Fourier transform of the basis 
$\{(\omega^{0})^{-r} (\omega^{1})^{-n-2+r}\}$ of 
$H^{1}(\mathbb{CP}^{1},\mathcal{O}(-n-2))$ is precisely the basis
$\{(\bar{\pi}_{0})^{r-1} (\bar{\pi}_{1})^{n+1-r}\}$ of
$H^{0}(\mathbb{CP}^{1},\mathcal{O}(n))$.
\end{proof}

We emphasize that
the mapping composed of the inverse Fourier transform and
the change of variables $(\bar{\pi}_{0},\bar{\pi}_{1}) \mapsto 
(\omega^{0},\omega^{1})$ is an isomorphism from 
$H^{1}(\mathbb{CP}^{1},\mathcal{O}(-n-2))$ to its dual space
$H^{0}(\mathbb{CP}^{1},\mathcal{O}(n))$.

\begin{corollary}
The momentum-space functions to which 
the Fourier transform is applied are represented as
\begin{align}
\bigoplus_{n=0}^{\infty} 
H^{0}(L,\mathcal{O}(n+2+m)) \otimes H^{0}(\mathbb{CP}^{1},\mathcal{O}(n)).
\label{30}
\end{align}
\end{corollary}

\begin{proof}
Let us consider subsets of the basis of Eq. $(\ref{29})$ that satisfy the
condition $j+k=n+2$ ($n=0,1,2,\cdots$). 
The elements of these subsets 
are products of $A_{jk}(\pi_{\dot{0}},\pi_{\dot{1}}) \in
H^{0}(L,\mathcal{O}(n+2+m))$ and  $1/(\omega^{0})^{j} 
(\omega^{1})^{k} \in H^{1}(\mathbb{CP}^{1},\mathcal{O}(-n-2))$, i.e.,
\begin{align}
&\left\{ \underset{j+k=n+2}{\sum_{j,k>0}} 
\dfrac{A_{jk}(\pi_{\dot{0}}, \pi_{\dot{1}})}{(\omega^{0})^{j} (\omega^{1})^{k}} 
\right\} 
\notag \\
&= H^{0}\left( L, \mathcal{O}(n+2+m)\right) \otimes 
H^{1}\left( \mathbb{CP}^{1}, \mathcal{O}(-n-2)\right).
\label{31}
\end{align}
Therefore, from Eqs. $(\ref{29})$ and $(\ref{31})$,
\begin{align}
&H^{1}\left( \mathbb{PT}-L, \mathcal{O}(m) \right)
\notag \\
&= \bigoplus_{n=0}^{\infty}
H^{0}\left(L, \mathcal{O}(n+2+m)\right) 
\otimes H^{1}\left(\mathbb{CP}^{1}, \mathcal{O}(-n-2)\right).
\label{32}
\end{align}
Because the inverse Fourier transform is an isomorphism from
$H^{1}(\mathbb{CP}^{1},\mathcal{O}(-n-2))$ to 
$H^{0}(\mathbb{CP}^{1},\mathcal{O}(n))$, taking the inverse Fourier
transform of Eq. $(\ref{32})$
 gives Eq. $(\ref{30})$.
\end{proof}

\section{Twistor operator representations}
\label{Trotto}

\subsection{Background}

In the quantization procedure in twistor theory, the variables
$\omega^{\alpha}$ and $\bar{\pi}_{\alpha}$ become the operators
$\hat{\omega}^{\alpha}$ and $\hat{\bar{\pi}}_{\alpha}$ satisfying
the commutation relations \cite{Pen5, PM, DN}
\begin{align}
\left[\hat{\omega}^{\alpha}, \hat{\bar{\pi}}_{\beta} \right] 
= \delta^{\alpha}_{\beta}, \quad
\left[\hat{\omega}^{\alpha}, \hat{\omega}^{\beta} \right] = 0, \quad
\left[\hat{\bar{\pi}}_{\alpha}, \hat{\bar{\pi}}_{\beta} \right] = 0.
\label{33}
\end{align}
By analogy with standard quantum mechanics, in the representation that
diagonalizes $\hat{\bar{\pi}}_{\alpha}$, these operators are represented as
\begin{align}
\hat{\bar{\pi}}_{\alpha} \doteq \bar{\pi}_{\alpha}, \quad
\hat{\omega}^{\alpha} \doteq \dfrac{\partial}{\partial \bar{\pi}_{\alpha}}.
\label{34}
\end{align}
Here, the symbol $\doteq$ stands for ``is represented by.''
In addition, in the representation that diagonalizes $\hat{\omega}^{\alpha}$,
\begin{align}
\hat{\bar{\pi}}_{\alpha} \doteq - \dfrac{\partial}{\partial \omega^{\alpha}},
\quad 
\hat{\omega}^{\alpha} \doteq \omega^{\alpha}.
\label{35}
\end{align}
Therefore, the following correspondence relations between 
elements $f(\bar{\pi}_{\alpha}, \pi_{\dot{\alpha}})$ of Eq. $(\ref{30})$ 
and their Fourier transforms 
$\mathcal{F}[f(\bar{\pi}_{\alpha}, \pi_{\dot{\alpha}})]
=\tilde{f}(\omega^{\alpha},\pi_{\dot{\alpha}})$ are desirable:
\begin{align}
\bar{\pi}_{\alpha} f(\bar{\pi}_{\alpha}, \pi_{\dot{\alpha}}) 
\leftrightarrow
- \dfrac{\partial}{\partial \omega^{\alpha}} 
\tilde{f}(\omega^{\alpha},\pi_{\dot{\alpha}}),
\label{36}
\end{align}
\begin{align}
\dfrac{\partial}{\partial \bar{\pi}_{\alpha}} 
f(\bar{\pi}_{\alpha}, \pi_{\dot{\alpha}}) 
\leftrightarrow
\omega^{\alpha} \tilde{f}(\omega^{\alpha},\pi_{\dot{\alpha}}).
\label{37}
\end{align}
In Sections \ref{4-1} and \ref{4-2}, 
we show that such correspondence relations exist, and
in Section \ref{4-3}, we identify the functions over which the operators
$\hat{\omega}^{\alpha}$ and $\hat{\bar{\pi}}_{\alpha}$ are closed.

\subsection{Fourier transform of the operators}
\label{4-1}

\begin{theorem}
The correspondence relations in Eqs. $(\ref{36})$ and
$(\ref{37})$ are satisfied by the Fourier transform, i.e.,
\begin{align}
\mathcal{F} \left[ \bar{\pi}_{\alpha} 
f(\bar{\pi}_{\alpha}, \pi_{\dot{\alpha}}) \right]
= - \dfrac{\partial}{\partial \omega^{\alpha}}
\mathcal{F} \left[ f(\bar{\pi}_{\alpha}, \pi_{\dot{\alpha}}) \right],
\label{38}
\end{align}
\begin{align}
\mathcal{F} \left[ \dfrac{\partial}{\partial \bar{\pi}_{\alpha}}
f(\bar{\pi}_{\alpha}, \pi_{\dot{\alpha}}) \right]
= \omega^{\alpha} \mathcal{F}
\left[ f(\bar{\pi}_{\alpha}, \pi_{\dot{\alpha}}) \right].
\label{39}
\end{align}
\end{theorem}

\begin{proof}
For the basis $\{(\bar{\pi}_{0})^{r-1} (\bar{\pi}_{1})^{n+1-r}\}_{r=1}^{n+1}$
of $H^{0}(\mathbb{CP}^{1},\mathcal{O}(n))$, it is seen 
from Eq. $(\ref{7})$
that
\begin{align}
\mathcal{F}\left[ \bar{\pi}_{0} (\bar{\pi}_{0})^{r-1} (\bar{\pi}_{1})^{n+1-r} \right]
= \dfrac{\textrm{sgn} \sigma}{(-2 \pi i)^{2}}
\dfrac{\varGamma (r+1) \varGamma (n+2-r)}
{(\omega^{0})^{r+1} (\omega^{1})^{n+2-r}},
\label{40}
\end{align}
\begin{align}
- \dfrac{\partial}{\partial \omega^{0}} 
\mathcal{F}\left[ (\bar{\pi}_{0})^{r-1} (\bar{\pi}_{1})^{n+1-r} \right]
= \dfrac{\textrm{sgn} \sigma}{(-2 \pi i)^{2}}
\dfrac{r \varGamma (r) \varGamma (n+2-r)}
{(\omega^{0})^{r+1} (\omega^{1})^{n+2-r}}.
\label{41}
\end{align}
Hence, we have
\begin{align}
\mathcal{F}\left[ \bar{\pi}_{0} (\bar{\pi}_{0})^{r-1} (\bar{\pi}_{1})^{n+1-r} \right]
= -\dfrac{\partial}{\partial \omega^{0}} \mathcal{F} \left[
(\bar{\pi}_{0})^{r-1} (\bar{\pi}_{1})^{n+1-r}
\right].
\label{42}
\end{align}
Furthermore, it is seen from Eq. $(\ref{7})$ that
\begin{align}
&\mathcal{F}\left[\dfrac{\partial}{\partial \bar{\pi}_{0}} 
\left(\bar{\pi}_{0}\right)^{r-1} \left(\bar{\pi}_{1}\right)^{n+1-r}\right]
\notag \\
&= \begin{cases}
   \dfrac{\textrm{sgn}\sigma}{(-2\pi i)^{2}} \dfrac{(r-1)\varGamma(r-1) \varGamma(n+2-r)}
   {\left(\omega^{0}\right)^{r-1} \left(\omega^{1}\right)^{n+2-r}} & (r \ne 1) \\
   0                                                                                 & (r=1)
   \end{cases},
\label{43}
\end{align}
\begin{align}
\omega^{0} \mathcal{F} 
\left[ (\bar{\pi}_{0})^{r-1} (\bar{\pi}_{1})^{n+1-r} \right]
= \dfrac{\textrm{sgn} \sigma}{(- 2 \pi i)^{2}}
\dfrac{\varGamma(r) \varGamma(n+2-r)}
{(\omega^{0})^{r-1} (\omega^{1})^{n+2-r}}. 
\label{44}
\end{align}
When $r=1$, Eq. $(\ref{44})$ becomes 
\begin{align}
\dfrac{\textrm{sgn} \sigma}{(-2 \pi i)^{2}}
\dfrac{\varGamma(n+1)}{(\omega^{1})^{n+1}}.
\label{45} 
\end{align}
This is equivalent to $0$ as the element of 
$H^{1}(\mathbb{CP}^{1},\mathcal{O}(-n-1))$. Hence, we have
\begin{align}
\mathcal{F} \left[ \dfrac{\partial}{\partial \bar{\pi}_{0}}
(\bar{\pi}_{0})^{r-1} (\bar{\pi}_{1})^{n+1-r} \right]
= \omega^{0} \mathcal{F} 
\left[ (\bar{\pi}_{0})^{r-1} (\bar{\pi}_{1})^{n+1-r} \right].
\label{46}
\end{align}
We can also obtain equations similar to 
Eqs. $(\ref{42})$ and $(\ref{46})$
for the variables $\bar{\pi}_{1}$ and $\omega^{1}$.
Hence, for the elements $f(\bar{\pi}_{\alpha}, \pi_{\dot{\alpha}})$
of Eq. $(\ref{30})$, we obtain Eqs. $(\ref{38})$ and 
$(\ref{39})$.
\end{proof}

\subsection{Inverse Fourier transform of the operators}
\label{4-2}

\begin{theorem}
The correspondence relations in Eqs. 
$(\ref{36})$ and $(\ref{37})$
are satisfied by the inverse Fourier transform, i.e.,
\begin{align}
\mathcal{F}^{-1} 
\left[ - \dfrac{\partial}{\partial \omega^{\alpha}}
\tilde{f}(\omega^{\alpha}, \pi_{\dot{\alpha}}) \right]
= \bar{\pi}_{\alpha} \mathcal{F}^{-1}
\left[ \tilde{f}(\omega^{\alpha},\pi_{{\dot{\alpha}}}) \right],
\label{47}
\end{align}
\begin{align}
\mathcal{F}^{-1} 
\left[ \omega^{\alpha} \tilde{f}(\omega^{\alpha},\pi_{\dot{\alpha}}) \right]
= \dfrac{\partial}{\partial \bar{\pi}_{\alpha}} \mathcal{F}^{-1}
\left[ \tilde{f}(\omega^{\alpha},\pi_{\dot{\alpha}}) \right].
\label{48}
\end{align}
\end{theorem}

\begin{proof}
For the basis $\{(\omega^{0})^{-r} (\omega^{1})^{-n-2+r}\}_{r=1}^{n+1}$
of $H^{1}(\mathbb{CP}^{1}, \mathcal{O}(-n-2))$, it is seen from 
Eq. $(\ref{16})$ that
\begin{align} 
\mathcal{F}^{-1} \left[ - \dfrac{\partial}{\partial \omega^{0}}
(\omega^{0})^{-r} (\omega^{1})^{-n-2+r} \right]
= (-2 \pi i)^{2} \textrm{sgn} \sigma 
\dfrac{r (\bar{\pi}_{0})^{r} (\bar{\pi}_{1})^{n+1-r}}
{\varGamma(r+1) \varGamma(n+2-r)},
\label{49}
\end{align}
\begin{align}
\bar{\pi}_{0} \mathcal{F}^{-1} 
\left[ (\omega^{0})^{-r} (\omega^{1})^{-n-2+r} \right]
= (-2 \pi i)^{2} \textrm{sgn} \sigma
\dfrac{(\bar{\pi}_{0})^{r} (\bar{\pi}_{1})^{n+1-r}}
{\varGamma(r) \varGamma(n+2-r)}.
\label{50}
\end{align}
Hence, we have
\begin{align}
\mathcal{F}^{-1} \left[ - \dfrac{\partial}{\partial \omega^{0}} 
(\omega^{0})^{-r} (\omega^{1})^{-n-2+r} \right]
= \bar{\pi}_{0} \mathcal{F}^{-1} 
\left[ (\omega^{0})^{-r} (\omega^{1})^{-n-2+r} \right].
\label{51}
\end{align}
Furthermore, it is seen from Eq. $(\ref{16})$ that
\begin{align}
&\mathcal{F}^{-1}\left[\omega^{0} \left(\omega^{0}\right)^{-r}
\left(\omega^{1}\right)^{-n-2+r}\right]
\notag \\
&= \begin{cases}
   (-2\pi i)^{2} \textrm{sgn} \sigma 
   \dfrac{\left(\bar{\pi}_{0}\right)^{r-2} \left(\bar{\pi}_{1}\right)^{n+1-r}}
   {\varGamma(r-1) \varGamma(n+2-r)}  & (r \ne 1) \\
   0                                                   & (r=1) 
   \end{cases},
\label{52}   
\end{align}
\begin{align}
&\dfrac{\partial}{\partial \bar{\pi}_{0}} \mathcal{F}^{-1} 
\left[\left(\omega^{0}\right)^{-r} \left(\omega^{1}\right)^{-n-2+r}\right]
\notag \\
&= (-2 \pi i)^{2} \textrm{sgn} \sigma 
\dfrac{(r-1)\left(\bar{\pi}_{0}\right)^{r-2} \left(\bar{\pi}_{1}\right)^{n+1-r}}
{\varGamma(r) \varGamma(n+2-r)}.
\label{53}
\end{align} 
Hence, we have
\begin{align}
\mathcal{F}^{-1} 
\left[ \omega^{0} (\omega^{0})^{-r} (\omega^{1})^{-n-2+r}\right]
= \dfrac{\partial}{\partial \bar{\pi}_{0}} \mathcal{F}^{-1}
\left[ (\omega^{0})^{-r} (\omega^{1})^{-n-2+r} \right].
\label{54}
\end{align}
We can also obtain equations similar to
Eqs. $(\ref{51})$ and $(\ref{54})$
for the variables $\omega^{1}$ and $\bar{\pi}_{1}$.
Hence, for the elements $\tilde{f}(\omega^{\alpha},\pi_{\dot{\alpha}})$ of
Eq. $(\ref{29})$, we obtain Eqs. $(\ref{47})$ and
$(\ref{48})$.
\end{proof}

\subsection{Applying the operators}
\label{4-3}

\begin{theorem}
The application of the operators 
$\hat{\bar{\pi}}_{\alpha} \doteq \bar{\pi}_{\alpha}$ and
$\hat{\omega}^{\alpha} \doteq \partial / \partial \bar{\pi}_{\alpha}$ is
closed on the  momentum-space functions
\begin{align}
\bigoplus_{m=-\infty}^{\infty} \bigoplus_{n=0}^{\infty}
H^{0} \left( L, \mathcal{O}(n+2+m) \right) \otimes
H^{0} \left( \mathbb{CP}^{1}, \mathcal{O}(n) \right),
\label{55}
\end{align}
to which the Fourier transform is applied.
\end{theorem}

\begin{proof}
By applying the multiplication operators $\hat{\bar{\pi}}_{\alpha}
\doteq \bar{\pi}_{\alpha}$ to the basis $\{(\bar{\pi}_{0})^{r-1}
(\bar{\pi}_{1})^{n+1-r}\}_{r=1}^{n+1}$ of $H^{0}(\mathbb{CP}^{1},
\mathcal{O}(n))$, for $n=0,1,2,\cdots$, we have
\begin{align}
\bar{\pi}_{\alpha} : H^{0} \left( \mathbb{CP}^{1}, \mathcal{O}(n)\right)
\rightarrow H^{0} \left( \mathbb{CP}^{1}, \mathcal{O}(n+1) \right).
\label{56}
\end{align}
In addition, applying
 the partial differential operator $\hat{\omega}^{0}
\doteq \partial / \partial \bar{\pi}_{0}$ gives
\begin{align}
\dfrac{\partial}{\partial \bar{\pi}_{0}} :
(\bar{\pi}_{0})^{r-1} (\bar{\pi}_{1})^{n+1-r} \mapsto
(r-1) (\bar{\pi}_{0})^{r-2} (\bar{\pi}_{1})^{n+1-r}.
\label{57}
\end{align}
When $r=1$, the result is $0$, and the same is true for the partial 
differential operator $\hat{\omega}^{1} \doteq \partial / \partial
\bar{\pi}_{1}$. Thus, for $n=0$, we have 
\begin{align}
\dfrac{\partial}{\partial \bar{\pi}_{\alpha}} :
H^{0} \left( \mathbb{CP}^{1}, \mathcal{O}(0) \right) \rightarrow 0,
\label{58}
\end{align}
and for $n=1,2,3,\cdots$, we have
\begin{align}
\dfrac{\partial}{\partial \bar{\pi}_{\alpha}} :
H^{0} \left( \mathbb{CP}^{1}, \mathcal{O}(n) \right) \rightarrow
H^{0} \left( \mathbb{CP}^{1}, \mathcal{O}(n-1) \right).
\label{59}
\end{align}
Thus, 
the application
of the operators is closed on Eq. $(\ref{55})$.
\end{proof}

\begin{theorem}
The application of the operators
$\hat{\bar{\pi}}_{\alpha} \doteq - \partial / \partial \omega^{\alpha}$
and $\hat{\omega}^{\alpha} \doteq \omega^{\alpha}$ is closed on
the  twistor-space functions
\begin{align}
&\bigoplus_{m=-\infty}^{\infty} H^{1} \left( \mathbb{PT} - L,
\mathcal{O}(m) \right)
\notag \\
=
&\bigoplus_{m=-\infty}^{\infty} \bigoplus_{n=0}^{\infty}
H^{0} \left( L, \mathcal{O}(n+2+m) \right)
\otimes
H^{1} \left( \mathbb{CP}^{1}, \mathcal{O}(-n-2) \right), 
\label{60}
\end{align}
to which the inverse Fourier transform is applied.
\end{theorem}

\begin{proof}
By applying the partial differential operators 
$\hat{\bar{\pi}}_{\alpha} \doteq -\partial /\partial \omega^{\alpha}$ 
to the basis $\{(\omega^{0})^{-r} (\omega^{1})^{-n-2+r}\}_{r=1}^{n+1}$
of $H^{1}(\mathbb{CP}^{1},\mathcal{O}(-n-2))$ 
, for $n=0,1,2,\cdots$, we have
\begin{align}
-\dfrac{\partial}{\partial \omega^{\alpha}} :
H^{1} \left( \mathbb{CP}^{1}, \mathcal{O}(-n-2) \right) \rightarrow
H^{1} \left( \mathbb{CP}^{1}, \mathcal{O}(-n-3) \right).
\label{61}
\end{align}
In addition, applying the multiplication operator
$\hat{\omega}^{0} \doteq \omega^{0}$ gives
\begin{align}
\omega^{0} : \left( \omega^{0} \right)^{-r} \left( \omega^{1} \right)^{-n-2+r}
\mapsto \left( \omega^{0} \right)^{-r+1} \left( \omega^{1} \right)^{-n-2+r}.
\label{62}
\end{align}
When $r=1$, we have $(\omega^{1})^{-n-1}$. This is equivalent to $0$
as the element of $H^{1}(\mathbb{CP}^{1}, \mathcal{O}(-n-1))$.
The same is true for the multiplication operator
$\hat{\omega}^{1} \doteq \omega^{1}$. Hence, for $n=0$, we have
\begin{align} 
\omega^{\alpha} : H^{1} \left( \mathbb{CP}^{1}, \mathcal{O}(-2) \right)
\rightarrow 0,
\label{63}
\end{align}
and  for $n=1,2,3,\cdots$, we have
\begin{align}
\omega^{\alpha} : H^{1} \left( \mathbb{CP}^{1}, \mathcal{O}(-n-2) \right)
\rightarrow
H^{1} \left( \mathbb{CP}^{1}, \mathcal{O}(-n-1) \right).
\label{64}
\end{align}
From this fact and Eq. $(\ref{32})$,
the application of the operators is closed on Eq. $(\ref{60})$.
\end{proof}

\section{Summary and discussion}
\label{Sd}
We proposed a new definition of the
 Fourier transform from momentum space to twistor space
as well as its inverse. Here, the momentum-space coordinates
are defined 
by the $SL(2,\mathbb{C})$ spinor $(\pi_{\dot{\alpha}})_{\dot{\alpha}=\dot{0},
\dot{1}}$ and its complex conjugate $(\bar{\pi}_{\alpha})_{\alpha=0,1}$, and the
twistor-space coordinates are defined by the $SL(2,\mathbb{C})$ spinors
$(\omega^{\alpha})_{\alpha=0,1}$ and 
$(\pi_{\dot{\alpha}})_{\dot{\alpha}=\dot{0}, \dot{1}}$.
First, we assumed that the momentum-space functions 
$f(\bar{\pi}_{\alpha}, \pi_{\dot{\alpha}})$ can be expanded 
in terms of the power series terms
$(\bar{\pi}_{0})^{a} (\bar{\pi}_{1})^{b}$ ($a,b=0,1,2,\cdots$).
Then,
we proposed the Fourier transform of $f(\bar{\pi}_{\alpha}, \pi_{\dot{\alpha}})$, $\mathcal{F}[f(\bar{\pi}_{\alpha}, \pi_{\dot{\alpha}})]$, as a complex
integral with respect to the variables $\bar{\pi}_{\alpha}$, which
 transforms $(\bar{\pi}_{0})^{0} (\bar{\pi}_{1})^{0}=1$
to the complex function of $\omega^{\alpha}$ representing the delta function
of Sato's hyperfunction.
By this definition, $\mathcal{F}[f(\bar{\pi}_{\alpha}, \pi_{\dot{\alpha}})]$
is the  twistor-space function 
$\tilde{f}(\omega^{\alpha},\pi_{\dot{\alpha}})$, which is expanded 
in terms of the
power series terms 
$(\omega^{0})^{-a-1}(\omega^{1})^{-b-1}$ ($a,b=0,1,2,\cdots$).
Next, we proposed the inverse Fourier transform of $\tilde{f}
(\omega^{\alpha},\pi_{\dot{\alpha}})$, 
$\mathcal{F}^{-1}[\tilde{f}(\omega^{\alpha},\pi_{\dot{\alpha}})]$,
as a complex integral that transforms the complex function of
$\omega^{\alpha}$ representing the delta function of Sato's hyperfunction
to 1.
Using this definition, we showed that
$\mathcal{F}^{-1}[\tilde{f}(\omega^{\alpha}, \pi_{\dot{\alpha}})]$
is precisely the inverse
map of $\mathcal{F}[f(\bar{\pi}_{\alpha},
\pi_{\dot{\alpha}})]$, i.e.,
$\mathcal{F}^{-1} \mathcal{F}[f(\bar{\pi}_{\alpha},\pi_{\dot{\alpha}})]
= f(\bar{\pi}_{\alpha},\pi_{\dot{\alpha}})$,
$\mathcal{F}\mathcal{F}^{-1}[\tilde{f}(\omega^{\alpha},\pi_{\dot{\alpha}})]
=\tilde{f}(\omega^{\alpha},\pi_{\dot{\alpha}})$.

To the best of our knowledge, this study is the first
to show
that this Fourier transform is an isomorphism
from $H^{0}(\mathbb{CP}^{1},\mathcal{O}(n))$ ($n=0,1,2,\cdots$)
to $H^{1}(\mathbb{CP}^{1},\mathcal{O}(-n-2))$, and that 
the inverse Fourier transform is an isomorphism from
$H^{1}(\mathbb{CP}^{1},\mathcal{O}(-n-2))$ to
$H^{0}(\mathbb{CP}^{1},\mathcal{O}(n))$, from the perspective of 
\v{C}ech cohomology groups.
In particular, owing to the Serre duality of the \v{C}ech cohomology groups,
the mapping consisting of the inverse Fourier transform and 
the change of variables $(\bar{\pi}_{0},\bar{\pi}_{0}) \mapsto
(\omega^{0},\omega^{1})$ is an isomorphism from
$H^{1}(\mathbb{CP}^{1},\mathcal{O}(-n-2))$ to its dual space
$H^{0}(\mathbb{CP}^{1},\mathcal{O}(n))$.

Further,  
we demonstrated
that the momentum-space functions 
to which  the Fourier transform is applied 
are $\oplus_{n=0}^{\infty} H^{0}(L,\mathcal{O}(n+2+m)) \otimes
H^{0}(\mathbb{CP}^{1},\mathcal{O}(n))$ ($m \in \mathbb{Z}$)
and that the twistor-space functions to which  the inverse Fourier transform 
is applied 
are
$H^{1}(\mathbb{PT}-L,\mathcal{O}(m))$.

We also showed that the representations of the operators 
$\hat{\bar{\pi}}_{\alpha}$ and $\hat{\omega}^{\alpha}$, which are 
well known in the twistor theory literature, are valid for this Fourier transform
and its inverse.
In addition, the application of these operators is closed on the
momentum-space functions $\oplus_{m=-\infty}^{\infty} \oplus_{n=0}^{\infty}
H^{0}(L,\mathcal{O}(n+2+m)) \otimes H^{0}(\mathbb{CP}^{1},\mathcal{O}(n))$
for the representation in which $\hat{\bar{\pi}}_{\alpha}$ reduces to
$\bar{\pi}_{\alpha}$, and on the twistor-space functions
$\oplus_{m=-\infty}^{\infty} H^{1}(\mathbb{PT}-L,\mathcal{O}(m))$
for the representation in which $\hat{\omega}^{\alpha}$ reduces to
$\omega^{\alpha}$.

{
In the real twistor space $\mathbb{RP}^{3}$, it is shown 
using Fourier transform that the three-particle MHV and
googly MHV super-amplitudes in $\mathcal{N}=4$ SYM,
which are the seed amplitudes for the twistor BCFW recursion, contain the so-called
sign factor \cite{MS2,ACCK2}.
This sign factor is the sign of the product of two twistors
and one infinity twistor; therefore, it breaks the conformal invariance.
Additionally,
the analogous factor also appears in the amplitudes in
$\mathcal{N}=8$ supergravity.
The origin of the sign factor is considered to be a
consequence of the fact that the amplitudes are 
functions on $\mathbb{RP}^{3}$.
In the complex twistor space $\mathbb{CP}^{3}$,
the amplitudes are elements of the \v{C}ech cohomology
group, rather than functions \cite{Pen6}.
If the amplitudes are treated as elements of the 
\v{C}ech cohomology
group on $\mathbb{CP}^{3}$, the sign factor will
not appear, because it will be incorporated within the
\v{C}ech cohomological structure \cite{MS2}.
Hence, in the future studies, we will reveal the origins of the sign
factor by studying the manner in which  
the sign factor is incorporated 
within the \v{C}ech cohomological structure of the 
Fourier transform formulated in this study.
}

We treated the twistor-space functions as
a \v{C}ech cohomology group.
However, in another formulation, they can also be
treated as a Dolbeault cohomology group \cite{WW}.
This formulation is in good agreement with twistor string theory.
Therefore, in the future, we will consider 
a concrete integral formulation of
the Fourier transform in terms of Dolbeault cohomology.

Finally, we note that the Fourier transform from momentum space 
to dual twistor space, i.e., the dual space of twistor space,
can be defined in the same manner as in this paper.

\section*{Acknowledgments}

I would like to thank S. Deguchi for his insightful comments.


\begin{thebibliography}{0}
\bibitem{Wit}
E. Witten, 
Perturbative gauge theory as a string theory
in twistor space,
{\it Commun. Math. Phys.} {\bf 252} (2004), 189-258; e-print arXiv:hep-th/0312171. 



\bibitem{ElHu}
H. Elvang and Y. -t. Huang,
Scattering amplitudes,
; e-print arXiv:1308.1697 [hep-th].\\
H. Elvang and Y. -t. Huang, {\it Scattering Amplitudes in Gauge Theory
and Gravity} (Cambridge University Press, Cambridge, 2015).


\bibitem{BBKR}
I. Bena, Z. Bern, D. A. Kosower and R. Roiban, 
Loops in twistor space,
{\it Phys. Rev.} {\bf D71} (2005), 106010; e-print arXiv:hep-th/0410054. 



\bibitem{BCF}
R. Britto, F. Cachazo and B. Feng,
New recursion relations for tree amplitudes of gluons,
{\it Nucl. Phys.} {\bf B715} (2005), 499-522; e-print
arXiv:hep-th/0412308. 



\bibitem{BCFW}
R. Britto, F. Cachazo, B. Feng and E. Witten,
Direct Proof of the Tree-Level Scattering Amplitude
Recursion Relation in Yang-Mills Theory,
{\it Phys. Rev. Lett.} {\bf 94} (2005), 181602; e-print
arXiv:hep-th/0501052.



\bibitem{MS2}
L. Mason and D. Skinner,
Scattering amplitudes and BCFW recursion in
twistor space,
{\it JHEP} {\bf 1001} (2010), 064; e-print arXiv:0903.2083
[hep-th].


\bibitem{ACCK2}
N. Arkani-Hamed, F. Cachazo, C. Cheung and J. Kaplan,
The S-matrix in twistor space,
{\it JHEP} {\bf 03}
(2010), 110; e-print arXiv:0903.2110 [hep-th].



\bibitem{ACCK}
N. Arkani-Hamed, F. Cachazo, C. Cheung and J. Kaplan,
A duality for the S matrix,
{\it JHEP} {\bf 1003}
(2010), 020; e-print arXiv:0907.5418 [hep-th].



\bibitem{MS}
L. Mason and D. Skinner,
Dual superconformal invariance, momentum twistors
and Grassmannians,
{\it JHEP} {\bf 0911} (2009), 045; e-print arXiv:0909.0250 [hep-th].



\bibitem{DNOS}
S. Deguchi, S. Negishi, S. Okano and T. Suzuki, 
Canonical formalism and quantization of a massless spinning bosonic particle
in four dimensions, {\it Int. J. Mod. Phys.} {\bf A29}
(2014), 1450044; e-print arXiv:1309.4169 [hep-th].




\bibitem{Pen1}
R. Penrose, Twistor algebra,
{\it J. Math. Phys.} {\bf 8} (1967), 345-366.


\bibitem{Pen5}
R. Penrose, Twistor quantisation and curved space-time,
{\it Int. J. Theor. Phys.} {\bf 1} (1968), 61-99. 


\bibitem{PM}
R. Penrose and M. A. H. MacCallum, 
Twistor theory: an approach to the
quantisation of fields and space-time,
{\it Phys. Rep.} {\bf 6} (1972), 241-316. 




\bibitem{DN}
S. Deguchi and J. Note,
(Pre-)Hilbert spaces in twistor quantization,
{\it J. Math. Phys.} {\bf 54} (2013), 072304; e-print arXiv:1210.0349 [hep-th].



\bibitem{Ka}
A. Kaneko, {\it Introduction to Hyperfunctions,} Mathematics and its
applications, Japanese series translated by Y. Yamamoto, 
edited by F. M. Arscott
(KTK Scientific Publishers, Kluwer Academic Publishers, Dordrecht, 1988).








\bibitem{HT}
S. A. Huggett and K. P. Tod, 
{\it An Introduction to Twistor Theory,} Second Edition, 
London Mathematical Society, Student Texts 4 
(Cambridge University Press, Cambridge, 1994). 



\bibitem{Koba}
R. O. Wells, Jr.,
{\it Differential Analysis on Complex Manifolds,} 3rd ed., Graduate Texts 
in Mathematics 65 (Springer-Verlag, New York, 2008).\\
S. Kobayashi, {\it Complex Geometry}
(Iwanami Shoten Co., Ltd., Tokyo, 2005) [in Japanese].





\bibitem{EH}
M. G. Eastwood and L. Hughston,
Massless fields based on a line: explosion and annihilation,
{\it Twistor Newsletter} {\bf 8} (1979), 43-47.


\bibitem{BE}
R. J. Baston and M. G. Eastwood, 
{\it The Penrose Transform: 
Its Interaction with Representation Theory,} 
Oxford Mathematical Monographs 499
(Clarendon Press, Oxford, 1989). 
\\
M. G. Eastwood and A. M. Pilato, 
On the density of twistor elementary states,
{\it Pacific J. Math.} {\bf 151} (1991), 201-215.




\bibitem{Pen6}
R. Penrose, Twistor functions and sheaf cohomology,
{\it Twistor Newsletter} {\bf 2} (1976), 3-12.



\bibitem{WW}
R. O. Wells, Jr., Complex manifolds and mathematical physics, {\it Bull. Amer. Math. Soc.} (new series)
{\bf 1} (1979), 296-336.
\\
N. M. J. Woodhouse, Twistor cohomology without sheaves, {\it Twistor Newsletter} {\bf 2} (1976), 13-14.

\end{thebibliography}
\end{document}